\documentstyle[11pt,twoside,colloqOHP2010,epsfig]{article}
\newsavebox{\leftbox}
\newlength{\leftboxheight}

\begin{document}
\title{The WASP-South search for transiting exoplanets}
\author{Coel Hellier$^{1}$, D.\,R.~Anderson$^{1}$, 
A.~Collier Cameron$^{2}$,  M.~Gillon$^{3}$, 
M.~Lendl$^{4}$,  T.\,A. Lister$^{5}$, P.\,F.\,L.~Maxted$^{1}$,  
D.~Pollacco$^{6}$,  D. Queloz$^{4}$, B.~Smalley$^{1}$,
A.\,H.\,M.\,J.~Triaud$^{4}$ and~R.\,G.~West$^{7}$}

\affil{$^{1}$Astrophysics Group, Keele University, UK\\ 
$^{2}$SUPA, Physics and Astronomy, University of St.\ Andrews, UK\\ 
$^{3}$Institut d'Astrophysique, Universit\'e de Li\`ege, Belgium\\ 
$^{4}$Observatoire astronomique de l'Universit\'e de Gen\`eve, CH\\ 
$^{5}$Las Cumbres Observatory, Santa Barbara, USA\\ 
$^{6}$Astrophysics Research Centre, Queen's University, Belfast, UK\\ 
$^{7}$Dept of Physics and Astronomy, University of Leicester, UK}

\begin{abstract}
Since 2006 WASP-South has been scanning the Southern
sky for transiting exoplanets. Combined with Geneva Observatory
radial velocities we have so far found over 30 transiting exoplanets
around relatively bright stars of magnitude 9--13. We present a 
status report for this ongoing survey. 
\end{abstract}
\section{WASP-South}
By 2004 several groups were pursuing systematic surveys
for transiting exoplanets, including HAT (Bakos et\,al.\ 2004),  OGLE
(Udalski et\,al.\ 2002), TrES (O'Donovan et\,al.\ 2006) and XO (McCullough
et\,al.\ 2006), and the
WASP consortium had begun operating a ``SuperWASP'' camera
array on La Palma in the Northern hemisphere (Pollacco et\,al.\ 2006).
In the South the OGLE project were finding planets around
stars of magnitude 15--16, but there was a clear
opportunity for WASP's wide-field, bright-sky strategy,
aimed at stars of magnitude 9--13.  

Funding from consortium universities was secured in 2004, leading to
the construction of WASP-South at the South African Astronomical
Observatory, closely copying Pollacco's design for the La Palma
SuperWASP.  Routine robotic operations commenced in March
2006. WASP-South is based around an array of 8 cameras, each with a
200-mm f/1.8 Canon lens backed by a {\it 2ev\/} 2048$\times$2048
Peltier-cooled chip. Each camera covers
7.8$^{\circ}\times$7.8$^{\circ}$, with 14$^{''}$ pixels, so that the
array covers 15$^{\circ}\times$30$^{\circ}$.  A broad-band filter
gives a 400--700 nm bandpass.

\begin{table}[t]
\caption{The WASP-South planets}
\begin{tabular}{lllllll}
\tableline
Name\rule{0mm}{3.5mm} & Position & Star & $V$ mag & Period & Mass & Radius \\ 
& & & & (days) & (Jup) & (Jup)  \\ [1mm] \tableline
WASP-4 \rule{0mm}{3.5mm}	&	2334--42 & 	G7V &	12.6 &	1.34 	&  1.12 &	1.42 \\ 
WASP-5 	&	2357--41 & 	G4V &	12.3 &	1.63 &	  1.64 &	1.17 \\ 
WASP-6 	&	2312--22 & 	G8 &	12.4 & 	3.36 &	  0.50 &	1.22 \\ 
WASP-7 	&	2044--39 & 	F5V & 	 {\ 9.5}& 	4.95& 	  0.96 &	0.92 \\ 
WASP-8 	&	2359--35 & 	G6 &	 {\ 9.9}& 	8.16& 	  2.24 &	1.04 \\ 
WASP-15 	&	1355--32 & 	F5 &	10.9& 	3.75& 	  0.54& 	1.43 \\ 
WASP-16 	&	1418--20 & 	G3V &	11.3& 	3.12& 	  0.86& 	1.01 \\ 
WASP-17 	&	1559--28 & 	F6 & 	11.6& 	3.74 &	  0.49 &	1.74 \\ 
WASP-18 	&	0137--45 & 	F9 & 	 {\ 9.3}& 	0.94& 	10.4& 	1.17 \\ 
WASP-19 	&	0953--45 & 	G8V & 	12.3& 	0.79& 	  1.15& 	1.31 \\ 
WASP-20 & & & 10.7 & 2.4	& \\
WASP-22 	&	0331--23 &  &		12.0 & 	3.53 &	  0.56&	1.12 \\ 
WASP-23 & 0644--42 & K1V & 12.7 & 2.94 & 0.87 & 0.96  \\
WASP-25 	&	1301--27 & 	G4 & 	11.9 & 	3.76 & 	  0.58	& 1.22 \\ 
WASP-26 	&	0018--15 & 	G0 & 	11.3 & 	2.76 & 	  1.02 &	1.32 \\ 
WASP-28 	&	2334--01 & 	F9 & 	12.0 & 	3.41 &	  0.91 &	1.12 \\ 
WASP-29 	&	2351--39 & 	K4V & 	11.3 & 	3.92 &	  0.24	&0.79 \\ 
WASP-30 	&	2353--10 & 	F8V &	11.9 &	4.16 &	61.0 &	0.89 \\ 
WASP-31 & 1117--19 & F7 & 11.7 & 3.41 & 0.47 & 1.56 \\ 
WASP-32 	&	0015+01 & &		11.3 &	2.72 & 	  3.60	& 1.18 \\ 
WASP-34 & 1101--23 & G5 & 10.4 & 4.32 & 0.59 & 1.22 \\
WASP-35 & & & 11 & 3.2  \\
WASP-36 & & & 12 & 1.5	& \\
WASP-37 	& 	1447+01  & 	G2 &	12.7 &	3.58 & 	  1.70 &	1.14 \\ 
WASP-40 & & & 12 & 3 & \\
WASP-41 & 1242--30 & G8V & 11.6 & 3.05 & 0.93 & 1.21 \\
\tableline
\end{tabular}
\end{table}

For the first two years WASP-South covered a ``zenith
strip'' centred at --32$^{\circ}$, but skipping fields
in the galactic plane that are too crowded.  We typically
raster 8--10 fields at a time with a cadence of $\sim$\,10 mins,
taking two 30-sec exposures at each pointing.  
For the next two years WASP-South covered an equatorial
strip centred at \mbox{--8$^{\circ}$}, partially overlapping 
with the Northern SuperWASP, while we are currently 
covering the far South with fields centred at --56$^{\circ}$. 

The data (approx 40 GB per night) are transferred on magnetic tape 
to Keele to be processed with the WASP pipeline. The data are 
then accumulated in an archive at Leicester where they are searched 
for candidate transits (see Collier-Cameron et\,al.\ 2006; 2007).  The main
limitations of the WASP design are the large pixels, often 
leading to blending, and the ``red noise'', which usually
exceeds photon noise at magnitudes brighter than 12.  Red noise
is introduced by changes in focus, and by drift of the images
over the chips through the night, by colour effects owing to
the wide bandpass, and by variations in sky
brightness such as through moonlight.  For this reason the
lightcurves are de-trended prior to the transit search.

\section{Planet discovery} 
Follow-up of WASP-South transit candidates involves
a major and ongoing collaboration with the Geneva Observatory, 
whose CORALIE spectrograph on the 1.2-m Euler telescope at 
La Silla is well matched to the $V$ = 9--13 brightness of 
WASP candidates.   Euler can also be switched rapidly to
a CCD camera, enabling us to obtain photometry
of candidates to confirm the reality of transits, and
particularly to check for possible blends with nearby stars.

Table~1 lists WASP-South planets as far as WASP-41. 
WASPs 35, 37 \&\ 40 were joint discoveries from combined
Southern and Northern WASP data, with radial-velocity
followup from OHP and the NOT in addition to CORALIE. 
Notable finds include the most bloated planet known and the first
found in a retrograde orbit, WASP-17b (Anderson et\,al.\ 2010a); 
the shortest-period transiting planet, WASP-19b (Hebb et\,al.\ 2010); 
the planet with the strongest known tidal interaction, WASP-18b
(Hellier et\,al.\ 2009); the highly eccentric 8-day system WASP-8b
(Queloz et\,al.\ 2010); and the transiting brown-dwarf 
WASP-30b (Anderson et\,al.\ 2010b).

Of the first 485 WASP candidates that were forwarded to 
Euler, 30 were found to be planets, 54 are still under
investigation, and 401 have been eliminated.  Thus our
overall hit rate is 1-in-14. The 
rejected candidates are overwhelmingly astrophysical
transit mimics, or blends with stars too close to be resolved in WASP
data.  Some stars prove to be too hot or fast rotating
for CORALIE radial velocities, while for
a small number of candidates the transit signal in the
WASP data has turned out to be a false alarm. 

\begin{figure}[t]
\includegraphics[width=78mm]{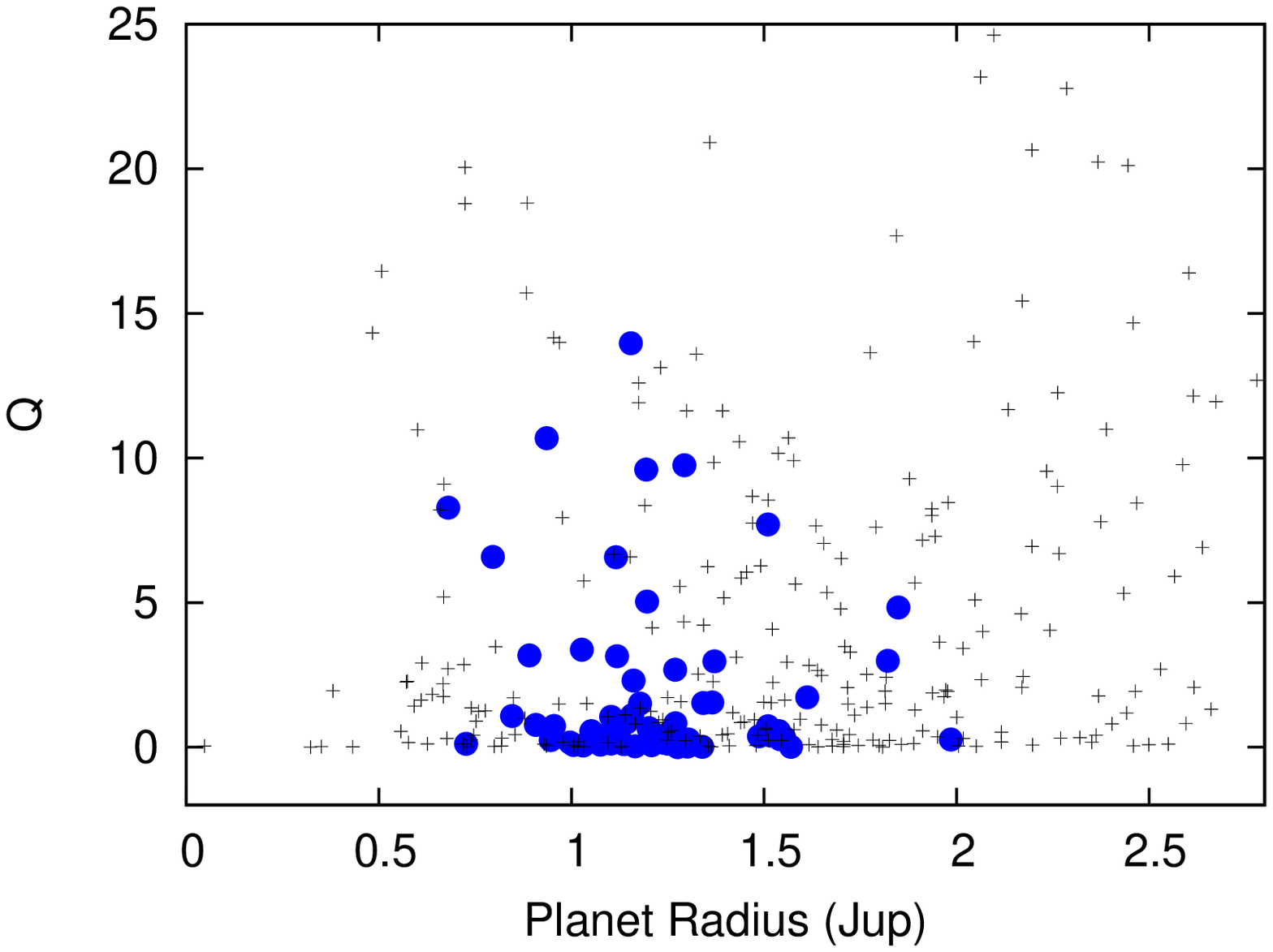}
\parbox[b]{4cm}{Figure~1: {\it Crosses are the companion radius and Q value 
for each candidate; circles are the actual planets.}}
\end{figure}

\begin{figure}[b]
\includegraphics[width=78mm]{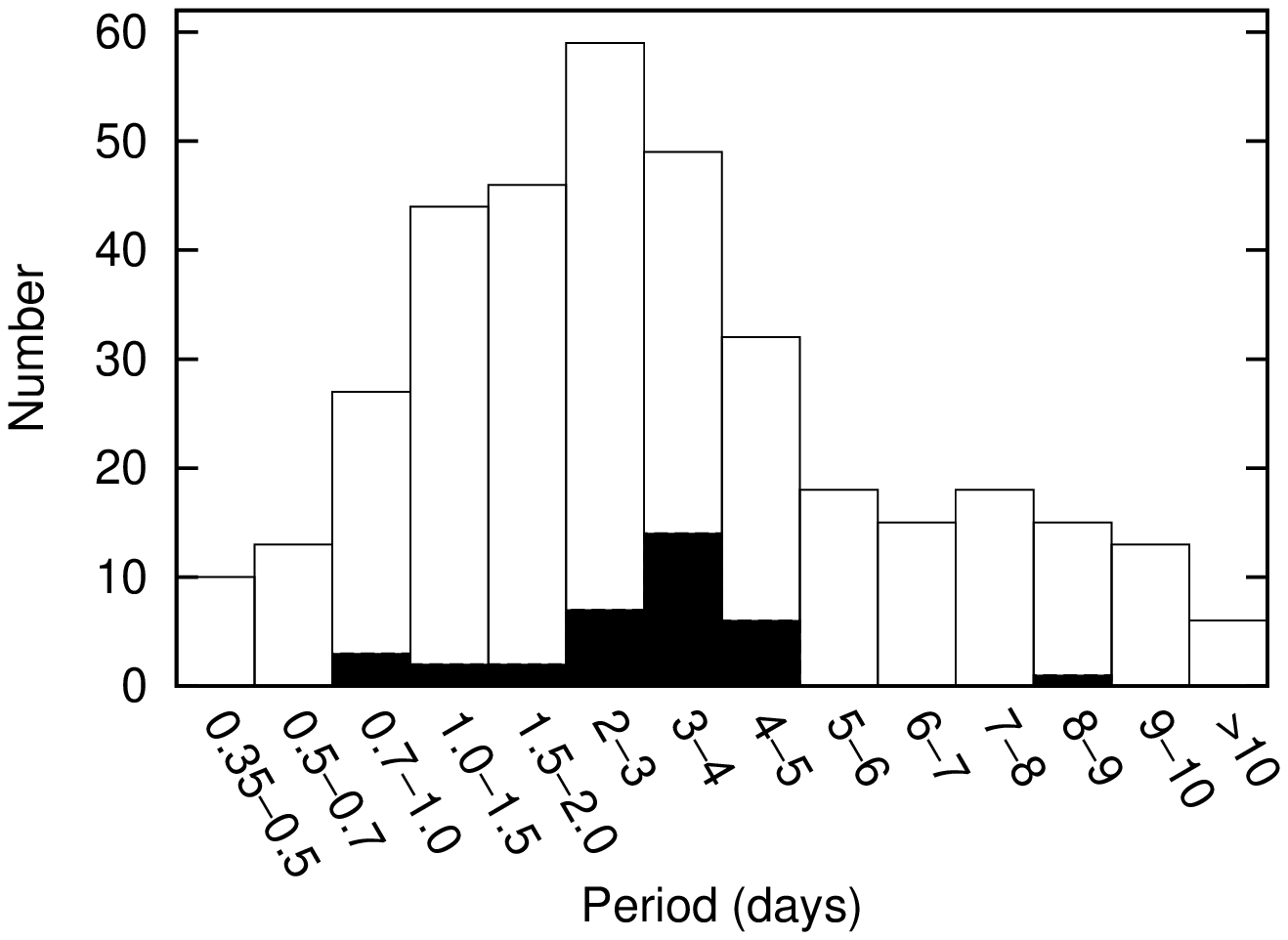}
\parbox[b]{4cm}{Figure~2: {\it The period distribution of WASP-South candidates
compared with that of the resulting planets.}}
\end{figure}

\addtocounter{figure}{2}

\section{Candidate selection} 
The selection of candidates has evolved over time, and
balances hit rate against the desirability of
finding unusual planets. For example Fig.~1 shows,
for each WASP-South candidate, the planet radius predicted 
from the WASP data and
the Q value, a measure of whether the
dip's width and depth is consistent with a planetary
transit around a main-sequence star (Collier-Cameron et\,al.\ 2007).  
Both values are only estimates, owing to the low S/N and
red noise of WASP data.

Candidates with 
predicted planet radii above 1.6 R$_{\rm Jup}$ are nearly 
all transit mimics, as expected, and so have a low hit rate. 
However, planets such as WASP-17b (Anderson et\,al.\ 2010a), with a 
radius near 2 R$_{\rm Jup}$, are among our most interesting
finds.   Candidates with large Q values are also poor
candidates, however a handful do turn out to be planets,
for example if the host star is evolved or if the WASP search 
algorithm has fitted a wrong impact parameter.

\begin{figure}[t]
\begin{center}
\epsfig{width=12cm,file=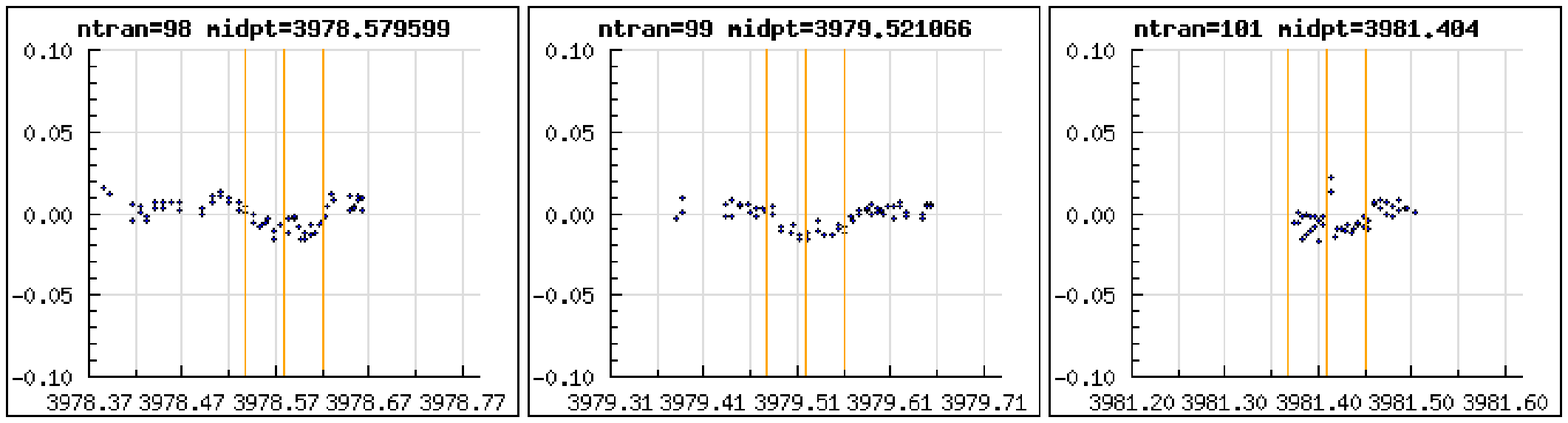}
\caption{Three of the sixteen discovery transits for WASP-18.}
\end{center}
\end{figure}

\begin{figure}[b]
\begin{center}
\epsfig{width=5.0cm,file=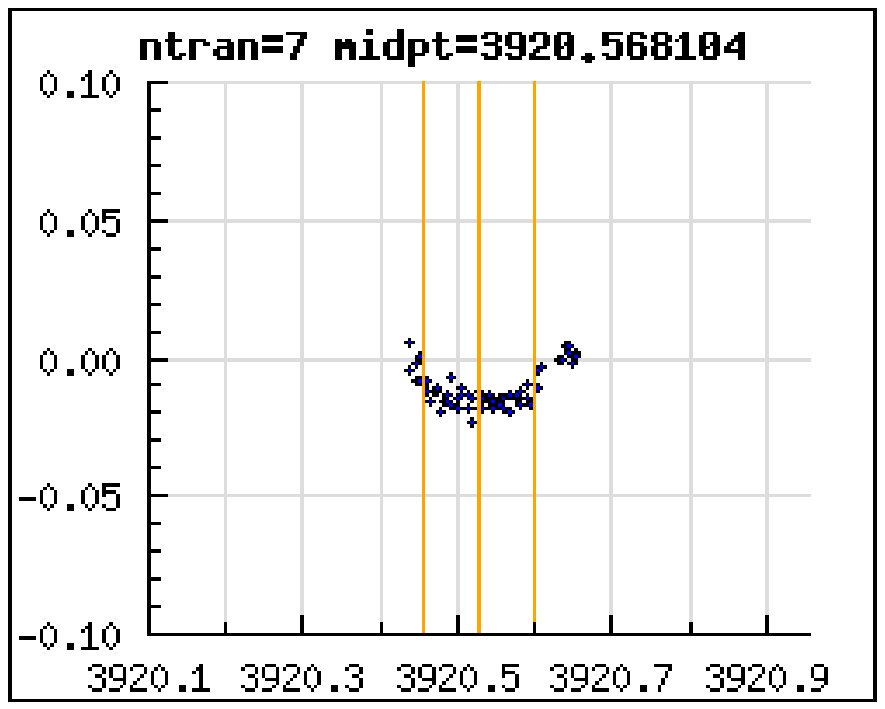}
\epsfig{width=5.0cm,file=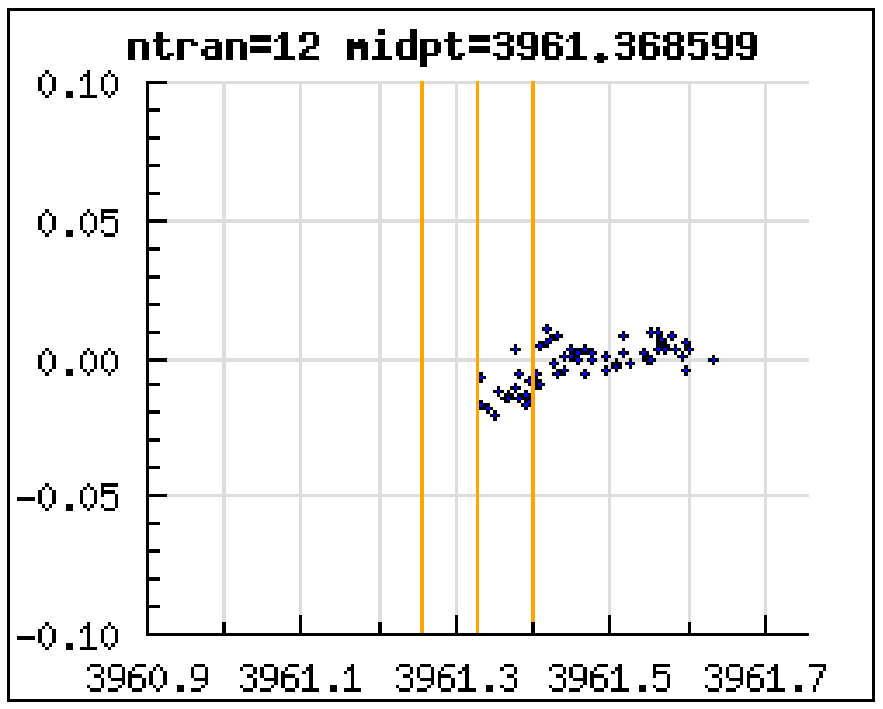}
\caption{The WASP discovery data for WASP-8.}
\end{center}
\end{figure}

Fig.~2 compares the period distribution of the WASP-South
candidates with the resulting planets, showing that most of
the planets that we find have periods of 2--5 days.  The lower
hit rate at periods 0.7--2.0 days is 
likely a real phenomenon, since these should be among the 
easiest planets for WASP to find (and note that HAT has similarly found 
only one planet below 2 d).   Above 5 days there are fewer
candidates, since the likelihood of covering several transits
of any object declines rapidly. The quality of the
candidates also declines, since with fewer transits we will be biased
towards deeper transits, which have a lower hit rate.

WASP-18b is an example of a planet that is relatively easy for 
WASP to find: with its short, 0.94-d
period, WASP-South covered 16 full or partial transits in
its first season, and at V = 9.3 the data quality is high (Fig.~3). 
In contrast, the discovery of the 8.16-d planet WASP-8b
was based on only one fairly full and one partial transit
(Fig.~4), but, given that it was early days
with few WASP-South candidates, that was considered
sufficient to justify CORALIE followup, even though the 
period was uncertain and only found from the radial-velocity
data. 

As another example we show in Fig.~5 the discovery 
data for the brown-dwarf system WASP-30,
which has the shallowest transit of the WASP-South
discoveries. Both the transit data and the accompanying
periodogram are highly marginal, which illustrates the
art of selecting candidates from the thousands 
thrown up by automatic search routines, especially
in the presence of red noise which can often mimic
transit features.  

\begin{figure}[tb]
\epsfig{width=7.5cm,file=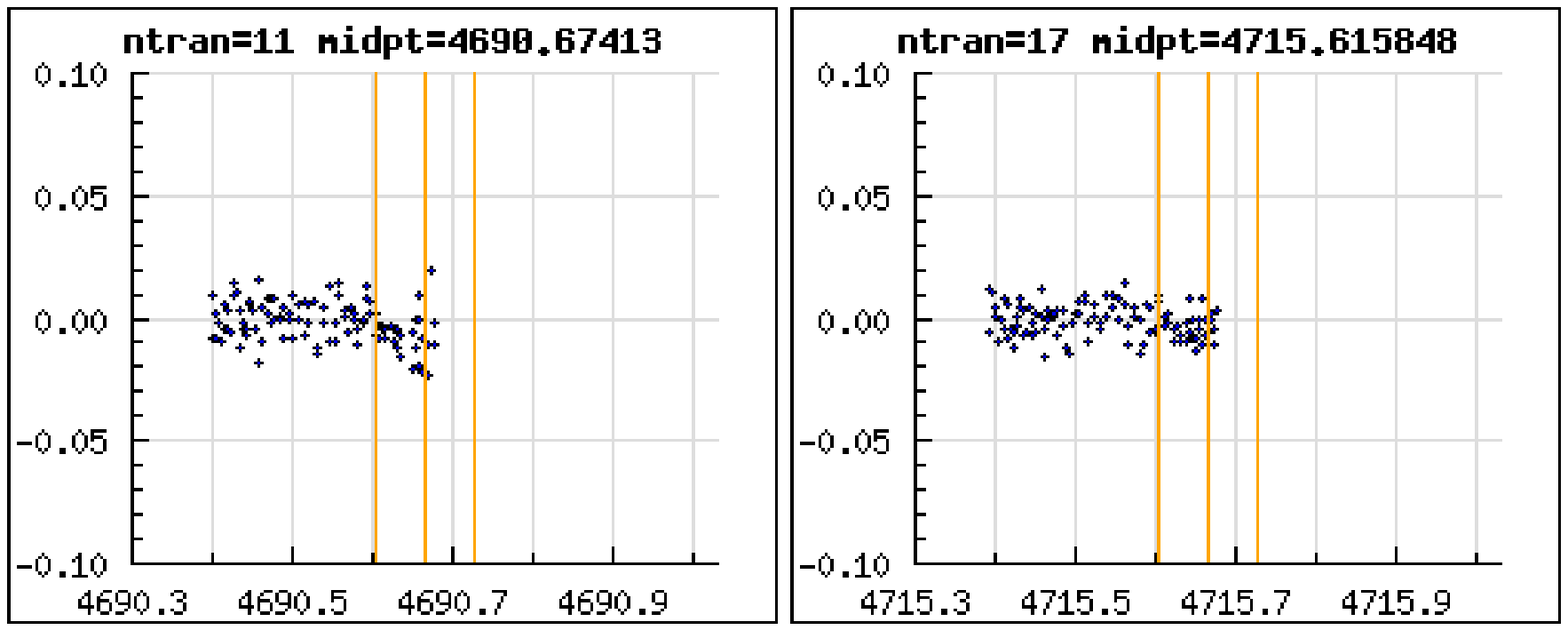}\\ [-6mm]
\epsfig{width=3.75cm,file=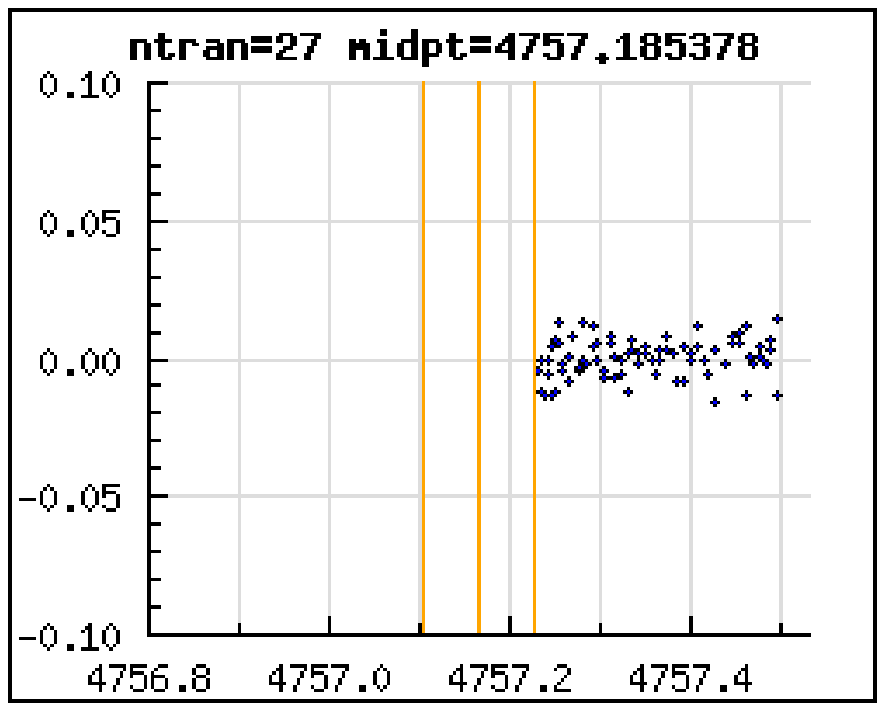}
\epsfig{width=3.75cm,file=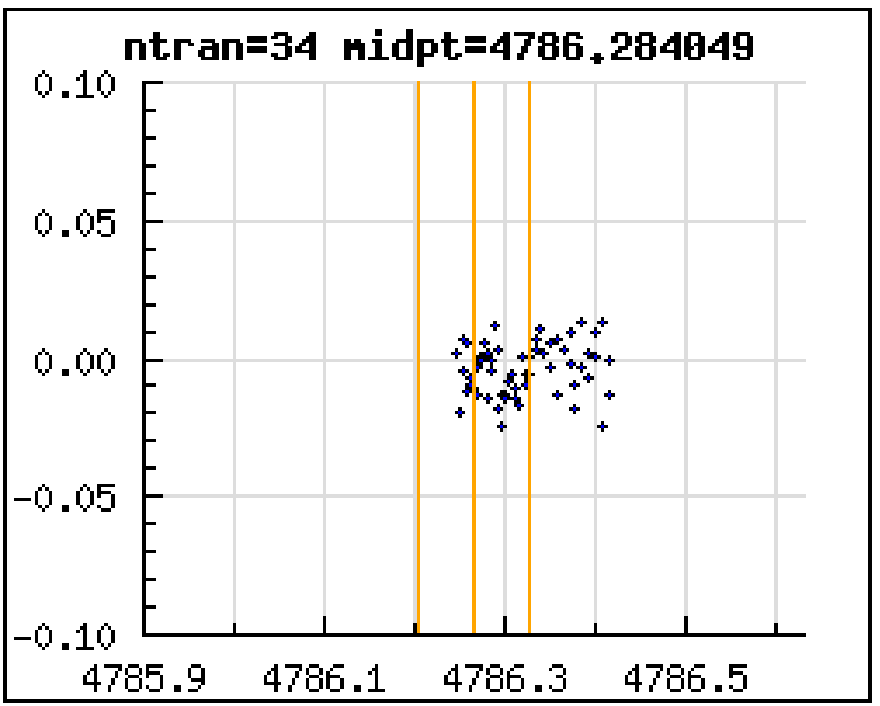}
\epsfig{width=4.0cm,file=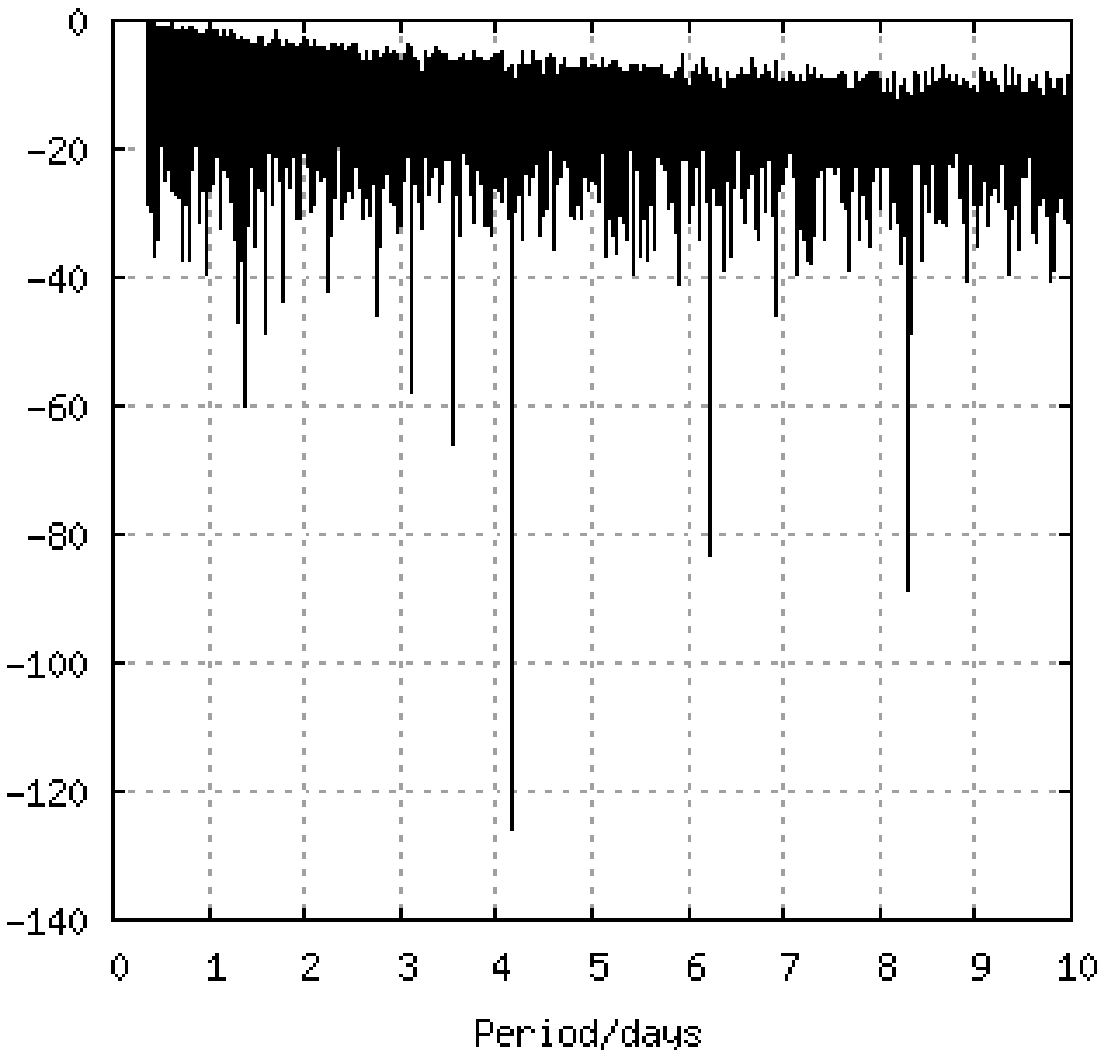}\\
\caption{WASP discovery transits and periodogram for WASP-30.}
\end{figure}

\begin{figure}[tb]
\begin{center}
\vspace{-1cm}\epsfig{width=12cm,file=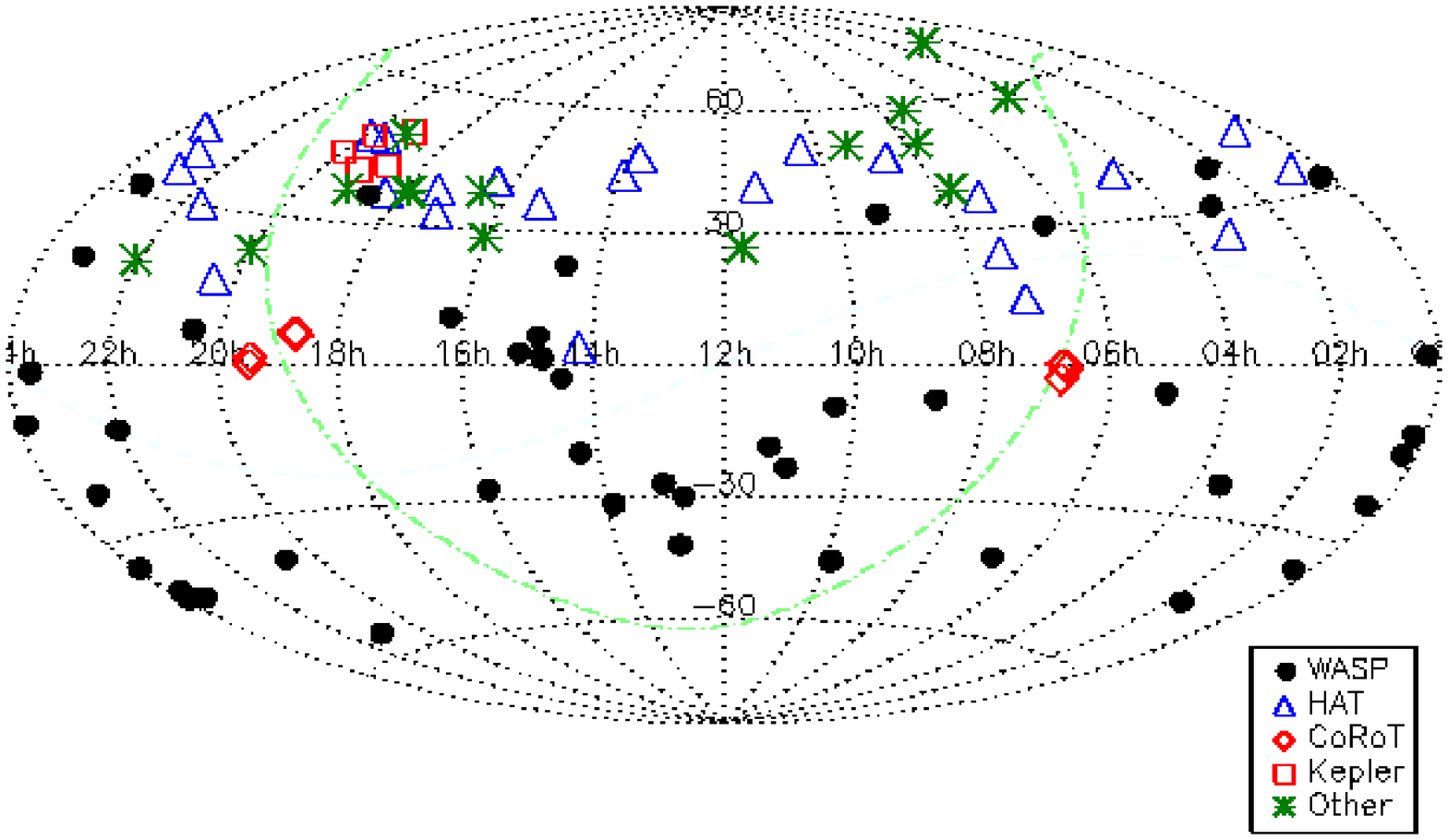}
\caption{Transiting planets around stars brighter than V = 14.}
\end{center}
\end{figure}

\section{WASP-South status}
WASP-South discoveries are ongoing, with 180 candidates
currently ``in play''.  
Clearly it would be possible to markedly improve our
hit-rate for finding Hot Jupiters, by using plots
such as Figs.~1 \&\ 2 to select optimum candidates, 
and by rejecting those with insecure transits or transit 
periods; however the interest will increasingly be 
not just in accumulating more 1-M$_{\rm Jup}$ systems in
4-day orbits, but in pushing the envelope of planet
parameters, especially by looking for smaller planets
that will produce the most marginal transit signals. 
Although we are running out of
virgin sky we expect to continue re-observing the previous
regions. Multiple seasons on the same regions of sky will
help to overcome red noise, pushing down to shallower 
transits, while also increasing the likelihood of finding 
longer-period systems.  

The distribution of known transiting 
exoplanets around bright stars (Fig.~6) shows that the
surveys are far from complete: the concentration
of known systems near the {\sl Kepler\/} field results
from the greater concentration of surveys on that region. 
Extrapolating to the rest of the sky implies that there 
are several hundred more transiting exoplanets within the discovery 
capabilities of WASP-like surveys.

\acknowledgments{We thank the South African Astronomical Observatory for hosting WASP-South and for their ongoing support and assistance.}


\begin{references}
\reference Anderson, D.R.  et\,al.\ 2010a, ApJ, 709, 159
\reference Anderson, D.R.  et\,al.\ 2010b, arXiv:1010.3006
\reference Bakos, G.A. et\,al.\ 2004, PASP, 116, 266
\reference Collier Cameron, A.  et\,al.\ 2006, MNRAS, 373, 799
\reference Collier Cameron, A.  et\,al.\ 2007, MNRAS, 380, 1230
\reference Hebb, L.   et\,al.\ 2007, 2010, ApJ, 708, 224
\reference Hellier, C.  et\,al.\ 2009 Nature, 460, 1098
\reference McCullough P.R. et\,al.\ 2006, ApJ, 648, 1228
\reference O'Donovan, F.T. et\,al.\ 2006, ApJ, 644, 1237 
\reference Pollacco, D.L. et\,al.\ 2006, PASP, 118, 1407
\reference Queloz, D.  et\,al.\ 2010, A\&A, 517, L1
\reference Udalski A. et\,al.\ 2002, Acta Astron., 52, 1 
\end{references}
\end{document}